\documentclass[useAMS,usenatbib]{mn2e} 

\usepackage{times}
\usepackage{graphicx}
\usepackage{amssymb}

\newcommand{\ion}[2]{{#1~\small#2}}

\newcommand{\HI}{$\rm H\,{\sevensize I}$}

\newcommand{\ha}{H$\alpha$}

\newcommand{\msun}{M$_\odot$}

\newcommand{\kms}{$\rm km~s^{-1}$}

\newcommand{\eso}{ESO137$-$001}

\title[A MUSE study of \eso]{MUSE sneaks a peek at extreme ram-pressure stripping events. I. A kinematic study of the archetypal galaxy \eso.}

\author[Fumagalli et al.]{Michele Fumagalli$^{1,2}$\thanks{E-mail: michele.fumagalli@durham.ac.uk}, 
  Matteo Fossati$^{3,4}$, George K.T. Hau$^{5}$, Giuseppe Gavazzi$^{6}$,\and
  Richard Bower$^{1}$, Ming Sun$^{7}$, Alessandro Boselli$^{8}$ \\
  $^{1}$Institute for Computational Cosmology, Department of Physics, Durham University, 
  South Road, Durham, DH1 3LE, UK \\
  $^{2}$Carnegie Observatories, 813 Santa Barbara Street, Pasadena, CA 91101, USA \\
  $^{3}$Universit{\"a}ts-Sternwarte M{\"u}nchen, Scheinerstrasse 1, D-81679 M{\"unchen}, Germany \\
  $^{4}$Max-Planck-Institut f{\"u}r Extraterrestrische Physik, Giessenbachstrasse, 
  D-85748 Garching, Germany\\
  $^{5}$European Southern Observatory, Alonso de Cordova 3107, Santiago, Chile \\
  $^{6}$Universit\`a di Milano-Bicoocca, Piazza della scienza 3, Milano, Italy \\
  $^{7}$Department of Physics, University of Alabama in Huntsville, Huntsville, AL 35899, USA \\
  $^{8}$Laboratoire d'Astrophysique de Marseille - LAM, Universit\'e d'Aix-Marseille \& CNRS, 
  UMR7326, 38 rue F. Joliot-Curie, F-13388 Marseille Cedex 13, France 
}

\begin{document}

\date{Accepted xxxx. Received xxxx; in original form xxxx}

\pagerange{\pageref{firstpage}--\pageref{lastpage}} \pubyear{xxxx}

\maketitle

\label{firstpage}

\begin{abstract}
We present MUSE observations of \eso, a spiral galaxy infalling towards the center of the 
massive Norma cluster at $z \sim 0.0162$. During the high-velocity encounter of \eso\ with the 
intracluster medium, a dramatic ram-pressure stripping event gives rise to an extended gaseous tail, traced
by our MUSE observations to $>30~$kpc from the galaxy center. By studying the \ha\ surface brightness 
and kinematics in tandem with the stellar velocity field, we conclude that ram pressure has completely 
removed the interstellar medium from the outer disk, while the primary tail is still fed by gas from 
the inner regions. Gravitational interactions do not appear to be a primary mechanism for gas removal. 
The stripped gas retains the imprint of the disk rotational velocity to $\sim 20~$kpc downstream,
without a significant gradient along the tail, which suggests that \eso\ is fast 
moving along a radial orbit in the plane of the sky. Conversely, 
beyond $\sim 20~$kpc, a greater degree of turbulence is seen, with velocity dispersion up to 
$\gtrsim 100~$\kms. For a model-dependent infall velocity of $v_{\rm inf} \sim 3000~$\kms, we conclude 
that the transition from laminar to turbulent flow in the tail occurs on timescales $\ge 6.5~$Myr. 
Our work demonstrates the terrific potential of MUSE for detailed studies of 
how ram-pressure stripping operates on small scales, providing a deep understanding of how galaxies 
interact with the dense plasma of the cluster environment.
\end{abstract}


\section{Introduction}

The discovery of a well-defined color bimodality in galaxy populations at redshifts $z = 0 - 2$
is arguably one of the most significant legacies of galaxy surveys. Extensive 
discussion on the color properties in large samples of galaxies 
can be found in the literature \citep[e.g.][]{bow92,str01,bal04,bel04,wei05,bra09}. 
A major task for modern studies of galaxy formation is therefore to understand how and when 
internal processes and environment drive the evolution of galaxies from the blue to the red cloud. 
While internal mechanisms, including ejective feedback from supernovae or active galactic nuclei,
are deemed responsible for suppressing and quenching star formation at all densities, 
environmental processes in the form of tidal interactions, harassment, and ram-pressure or viscous 
stripping are key players in shaping the observed color bimodality within rich galaxy clusters, 
especially at the faint end of the galaxy luminosity function. 
The origin and effects of these processes have been the subject of scrutiny of many
studies and review articles 
\citep[e.g.][]{lar80,has98,lew02,gav02,kau04,balo04,bos06,bos08,gav10,smi12,dre13}.

\begin{figure*}
\includegraphics[scale=0.58]{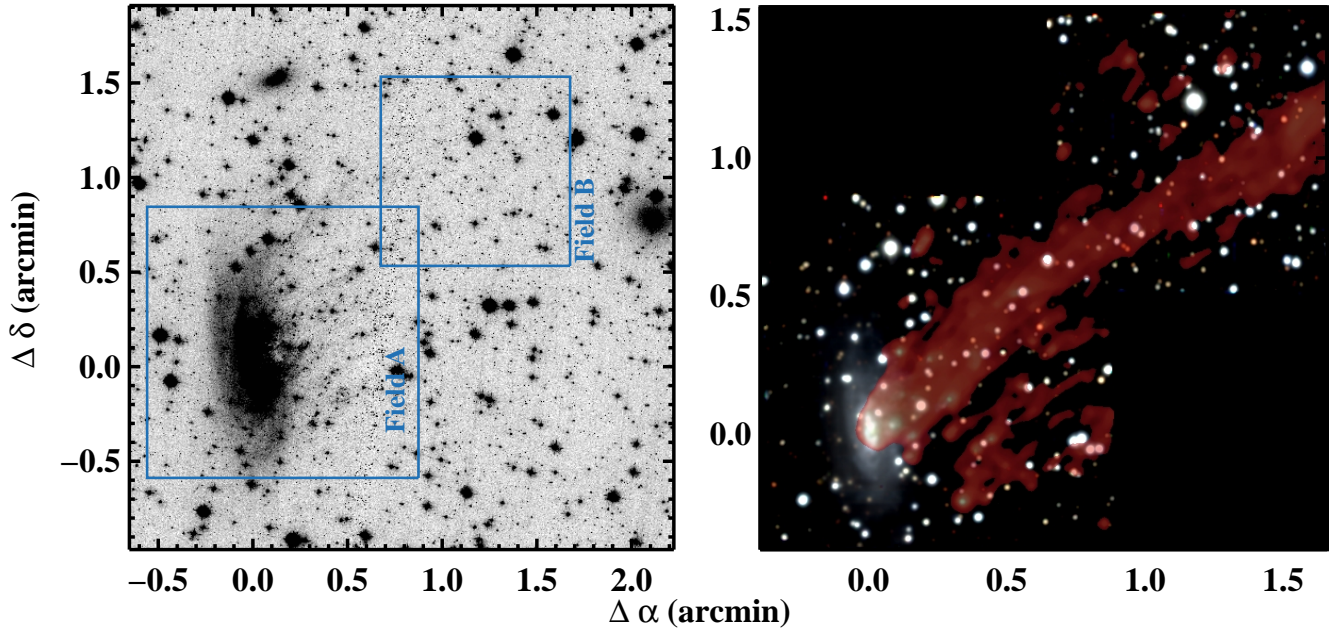}
\caption{{\it Left:} Archival HST/ACS image in the F475W filter (PID 11683; PI Ming Sun) 
with, superposed, the MUSE field of view at the two locations targeted by our observations. 
In this figure, north is up and east is to the left. The coordinate system is centered at the J2000 
position of \eso\ ($\alpha=$16:13:27.3, $\delta=$-60:45:51). At the redshift of \eso, 
$1' = 18.4$ kpc. {\it Right:} RGB color image obtained combining 
images extracted from the MUSE datacube in three wavelength intervals ($\lambda = 5000-6000~$\AA\ for the 
B channel, $\lambda = 6000-7000~$\AA\ for the G channel, and $\lambda = 7000-8000~$\AA\ for the R channel). 
A map of the \ha\ flux is overlaid in red using a logarithmic scale, revealing the extended gas
tail that originates from the high-velocity encounter of \eso\ with the intracluster medium.}\label{fig:fov}
\end{figure*}

\begin{figure*}
\includegraphics[scale=0.55]{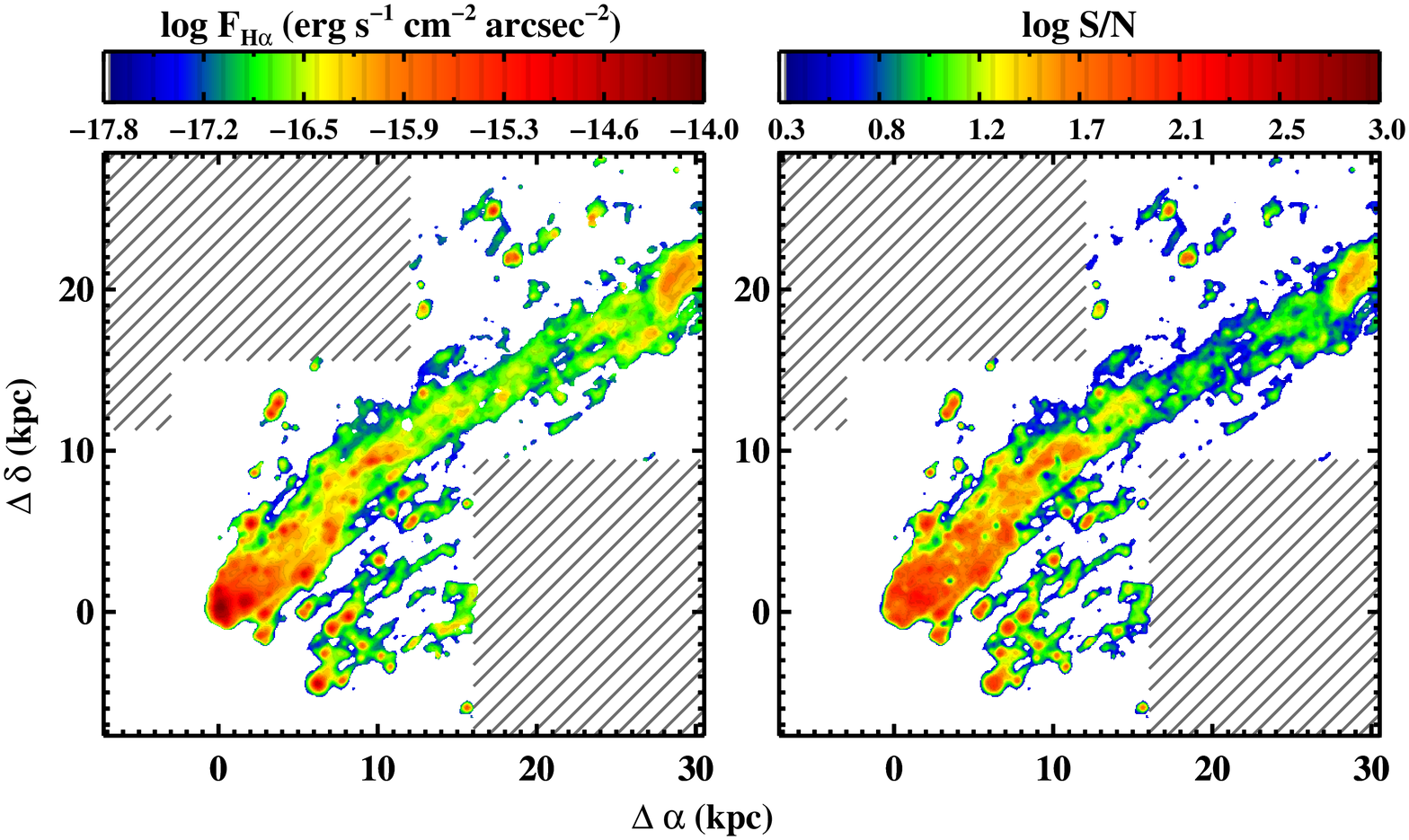}
\caption{{\it Left:} Map of the \ha\ surface brightness at a resolution of 
$60\times60$ pc, obtained from the MUSE datacube as described in the text.
Fluxes have been corrected for foreground Galactic extinction, but not for internal 
dust extinction. For visualization purposes, the map has been convolved with a median filter of $5\times 5$ 
pixels. {\it Right:} The $S/N$ map obtained by dividing the flux values by the associated 
errors in each pixel of the filtered datacube, which we have rescaled as described in the text. 
The dashed regions mark areas that are not covered by our observations. The image coordinate system 
is the same as the one adopted in Figure \ref{fig:fov}, but in proper physical units.
The unprecedented depth of our MUSE observations uncover an extended primary tail, as well as
a secondary southern tail with embedded HII regions.}\label{fig:flux}
\end{figure*}

The smoking gun of ongoing environmental transformation in nearby clusters is the disturbed 
gas content of member galaxies. Globally, radio surveys at 21 cm reveal that spiral galaxies in rich environments 
are deficient of \HI\ gas compared to matched samples at lower densities \citep[e.g.][]{hay84,gio85}. 
A deficiency of molecular gas has also been reported \citep{fum09,bos14}.
More significantly, these gas-deficient galaxies, and particularly those residing near the cluster cores, 
often exhibit truncated and disturbed gas disks \citep[e.g.][]{vol08,chu09} or, in some cases,
one-sided \HI\ tails \citep[e.g.][]{chu07}. 
Furthermore, extreme cases of disturbance are seen in ``head-tail'' radio continuum 
emission or in prominent \ha\ and UV tails arising from galaxies moving towards the centers of 
massive clusters, such as Coma, Virgo, and A1367 
\citep[e.g.][]{gav85,gav01,yos02,cor06,ken08,smi10,yag10,fos12}.

Detailed multiwavelength studies of these archetypal galaxies which are 
undergoing extreme gravitational and hydrodynamic transformations in the harsh cluster environment 
become an invaluable tool for unveiling the rich physics responsible for the morphological and color 
transformation of cluster galaxies. Many examples of these peculiar objects have been reported in
the recent literature \citep[e.g.][]{vol04,sun06,fum08,hes10,fum11,arr12,fos12,ken14,ebe14}.
In synergy with models and numerical simulations 
\citep[e.g.][]{gun72,nul82,moo96,qui00,sch01,vol01,bek03,kro08,ton10,rus14}, these
peculiar systems are in fact ideal laboratories to constrain the efficiency with which 
gas can be removed, quenching star formation. Extrapolated to the more general cluster population, 
results from these studies offer a unique perspective for how the red sequence is assembled 
in dense environments.

A particularly stunning example of a galaxy which is undergoing environmental 
transformation is \eso, member of the massive Norma cluster.
The Norma cluster, also known as A3627, has a dynamical mass of 
$M_{\rm dyn,cl} \sim 1\times 10^{15}~$\msun\ and lies in the Great Attractor region 
at a radial velocity of $v_{\rm cl}= 4871 \pm 54~$\kms, or redshift 
$z_{\rm cl}=0.01625\pm 0.00018$ \citep{wou08,nis12}. 
\eso\ is located at a projected distance of 
only $14.5'$ from the central cluster galaxy WKK6269, corresponding to $\sim 267$ kpc for the 
adopted cosmology \citep[$H_0=69.7$ and $\Omega_{\rm m}=0.236$ for which $1' = 18.4$ kpc;][]{hin13}.
Observations across the entire electromagnetic spectrum, including X-ray, optical, infrared, and  
millimetric data \citep{sun06,sun07,sun10,siv10,jac14}, show that \eso\ is suffering from 
an extreme ram-pressure stripping event during its first approach to the center of A3627.
Most notably, \eso\ exhibits a double tail which extend for $\sim 80$ kpc as seen in X-ray, 
pointing in the opposite direction from the cluster center \citep{sun06,sun10}. This tail, which is 
also detected in \ha\ \citep{sun07} and contains both cold and warm molecular gas \citep{siv10,jac14},
is thought to originate from the interaction between the hot intracluster medium (ICM) and the 
warm/cold interstellar medium (ISM) which is being ablated from the stellar disk by ram pressure
\citep[e.g.][]{sun10,jac14}.

A complete picture of how ram pressure operates in this galaxy and how the cold, warm, and hot gas 
phases coexist and interact inside the extended tail requires a detailed knowledge of the gas kinematics 
and of emission-line ratios. To date, these diagnostics have been limited to pointed radio observations
\citep{jac14} or spectroscopy of selected HII regions \citep{sun07,sun10}. 
To dramatically improve the status of current observations, we exploit the unprecedented capabilities 
the Multi Unit Spectroscopic Explorer \citep[MUSE;][]{bac10} at the UT4 Very Large Telescope.
With its unique 
combination of high efficiency ($\sim 0.35$ at 6800~\AA), extended wavelength coverage 
($\sim 4800-9300~$\AA\ sampled at $1.25~$\AA), and large field of view ($1'\times 1'$ in Wide Field Mode, 
sampled with $0.2''$ pixels), MUSE is a unique instrument 
to map at optical wavelengths the kinematics and line ratios of the disk and tail of \eso.
In this paper, we present MUSE observations of \eso\ which were conducted as part of the 
science verification 
run. After discussing the data reduction (Sect. \ref{sec:obs}) and the data analysis technique 
(Sect. \ref{sec:ana}), we present a study of the kinematics of the \ha\ emitting gas and of the 
stellar disk of \eso\ (Sect. \ref{sec:disc}). A subsequent paper in this series 
(Fossati et al., in prep.) 
will analyze emission and absorption line ratios to characterize the physical properties of the 
emitting gas and of the stellar populations in the disk and tail of \eso.

\begin{figure*}
\includegraphics[scale=0.55]{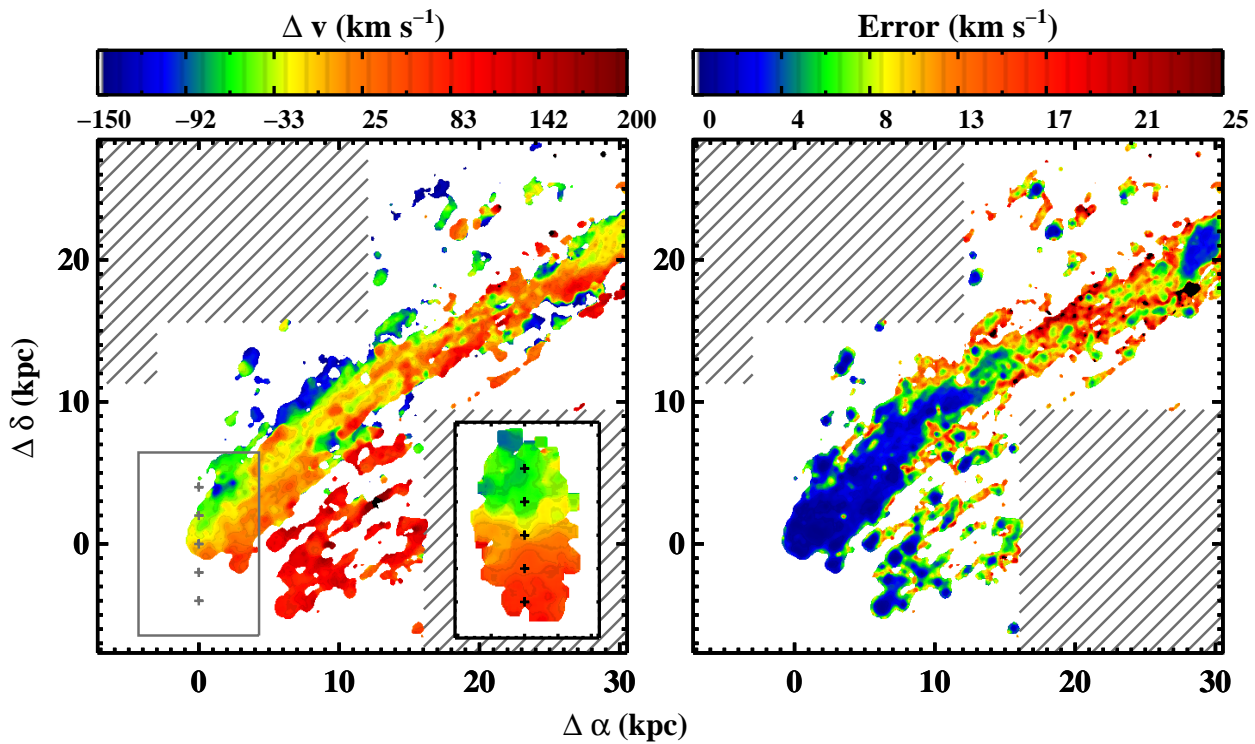}
\caption{{\it Left:} Velocity map of the \ha\ line relative to the galaxy systemic redshift
at a resolution of $60\times 60$ pc, obtained from the MUSE datacube as described in the text. 
For visualization purposes, the map has been convolved with a median filter of $5\times 5$ pixels. 
In the inset, we show the stellar kinematics from Figure \ref{fig:stkin}, on the same color 
scale adopted for the gas velocity map. To facilitate the comparison, we show the position of the
inset in the main panel (grey box), and we use crosses as a ruler with steps of 2 kpc. 
{\it Right:} The error map of the fitted centroids, rescaled as described 
in the text. The dashed regions mark areas which are not covered by our observations. 
The image coordinate system is the same as the one adopted in Figure \ref{fig:flux}.
The gas and stellar kinematics are remarkably aligned to distances of $\sim 20~$kpc 
along the tail, suggesting that \eso\ is fast moving along a radial orbit in the plane of 
the sky.}\label{fig:vel}
\end{figure*}

\begin{figure*}
\includegraphics[scale=0.55]{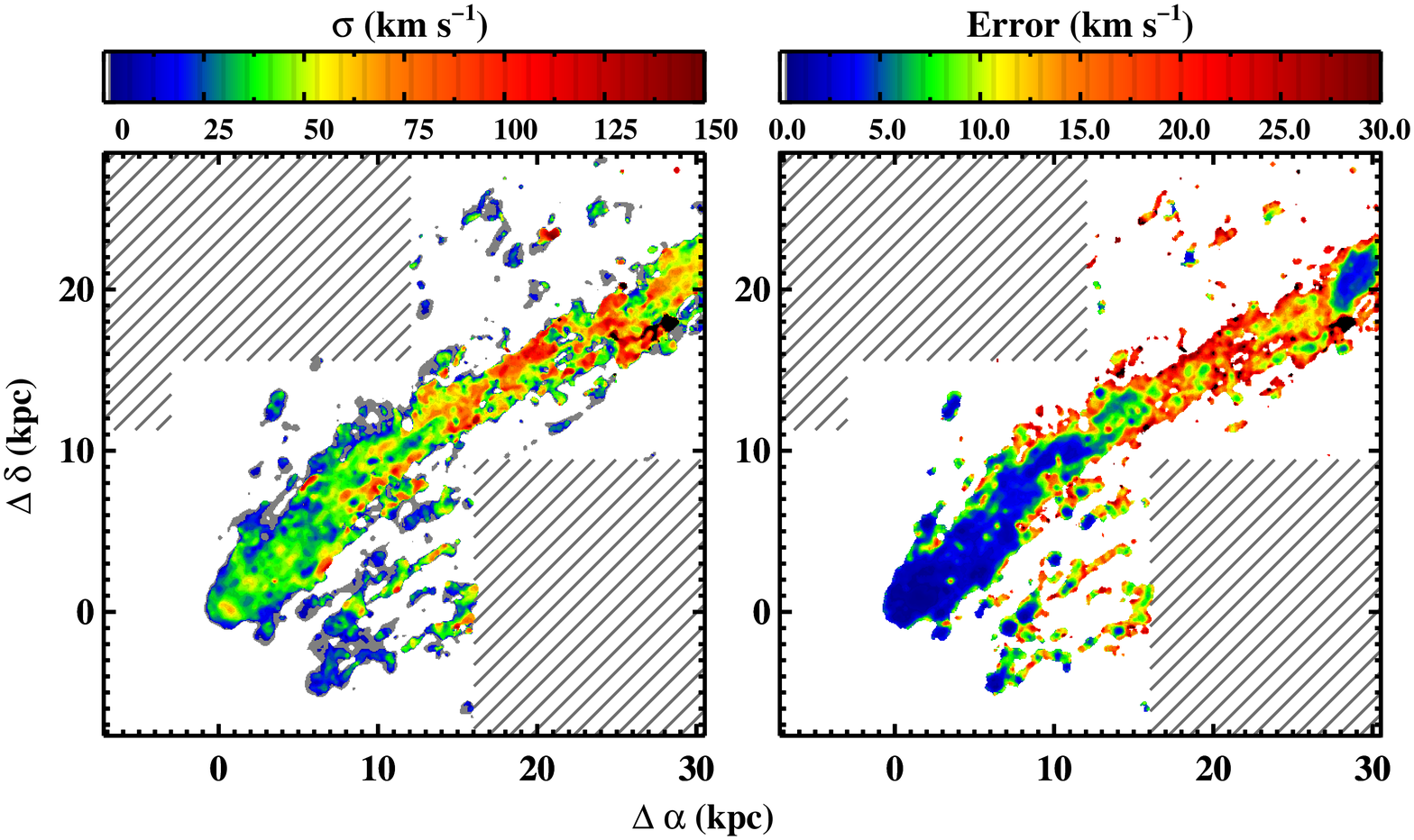}
\caption{{\it Left:}  Velocity dispersion map of the \ha\ line at a 
resolution of $60\times 60$ pc, obtained from the MUSE datacube as described in the text. 
Values have been corrected for instrumental resolution, and pixels in which lines are 
unresolved are masked in grey in the online edition. For visualization purposes, the map has been convolved with a 
median filter of $5\times 5$ pixels. {\it Right:} The error map of the fitted line widths, 
rescaled as described in the text. The dashed regions mark areas which are not covered by our observations.
The image coordinate system is the same as the one adopted in Figure \ref{fig:flux}. The transition 
from laminar to turbulent flow is seen at $\sim 20~$kpc from the galaxy center. 
For a model-dependent velocity of $\sim 3000~$\kms, turbulence arises on timescales 
$\ge 6.5$Myr.}\label{fig:sig}
\end{figure*}

\section{Observations and data reduction}\label{sec:obs}

\eso\ was observed with MUSE during the science verification run on 
UT June 21, 2014, under program 60.A-9349(A) (PI Fumagalli, Hau, Slezak).
Observations were conducted while the galaxy was transiting at 
airmass $\sim 1.25$, in photometric conditions and good seeing 
($\sim 0.7''-0.8''$). The disk and the inner tail of \eso\ (``Field A'' in Figure \ref{fig:fov}) 
were observed with three exposures of 900s each, with dithers of $13''-16''$ 
and a rotation of 90 degrees in the instrument position angle at each position.
A single 900s exposure (``Field B'' in Figure \ref{fig:fov}) was then acquired to cover 
the extent of the primary tail. Natural seeing mode and slow guiding mode were used.
 
The final data product is obtained using a combination of recipes from the 
early-release MUSE pipeline (v0.18.1) and a set of in-house IDL codes, which we have developed 
to improve the quality of the illumination, sky subtraction, and flux calibration of the MUSE data
specifically for the problem at hand. First, using the MUSE pipeline, we construct for each 
IFU a master bias, a master flat field, 
and a master sky flat, together with a wavelength solution. Next, we apply these 
calibrations to each science exposure and to exposures of the standard star GD153, 
which was observed for 30s at the beginning of the night in photometric conditions. 
After computing the instrument response curve for flux calibration and the telluric correction spectrum 
within the MUSE pipeline, we apply a smoothing kernel of 80 pixel to the 
obtained sensitivity function so as to remove residual small-scale fluctuations in the 
instrument response, which are present in the current version of the MUSE pipeline. 
The quality of photometric calibration is confirmed by aperture photometry 
on the $r$-band image reconstructed from the final datacube, which is in excellent agreement with 
$r$-band magnitudes in APASS, the AAVSO All-Sky Photometric Survey \citep{hen12}.

Using the MUSE pipeline, we then produce resampled datacubes for each exposure (in spaxels of
1.25\AA\ and $0.2''\times 0.2''$), also saving individual calibrated pixel tables which contain the data 
and variance pixels before interpolation. Cosmic rays are also identified as 7$\sigma$ outliers. At this stage, 
the MUSE pipeline could be used to compute the sky subtraction with the algorithm offered by the 
pipeline recipes. However, when sky subtraction is 
enabled, the reconstructed cube exhibits an evident variation in the illumination of each slice which, 
as noted in the pipeline manual, is currently not fully corrected by sky flats. 
Due to this variation, and perhaps to the particularly crowded nature of our field, the 
sky-subtracted cubes exhibit large negative residuals which would affect our subsequent analysis. 
To improve the data quality, we perform both an illumination correction and sky subtraction, 
using a set of IDL procedures which we have specifically developed to handle the reduced MUSE pixel tables. 
First, the illumination correction is computed by fitting eight bright sky lines in all slices 
within each exposure and by normalizing them to the median flux distribution across the entire field of view. 
After applying this illumination correction, we then generate a sky model in each exposure by combining pixels 
of comparable wavelength and within the 20-30th flux percentiles in a master sky spectrum. 
This operation is performed on the pixel table, to improve the spectral sampling of the sky model. 
We then subtract the sky model from each pixel, by interpolating the master sky spectrum 
with a spline function. Following these procedures, we find that the newly reconstructed datacubes present 
a more uniform illumination pattern and do not have evident residuals in the background level. 
As a further test of our procedure, we verify that the line flux from extragalactic sources 
in the sky-subtracted cubes and the non-sky subtracted cubes agree to within the typical flux 
standard deviation (see below). At each step, our IDL procedures propagate errors in the datacube 
which contains the pixel variance (also known as ``stat'' datacube).

At the end of the reduction, we combine the four exposures and we reconstruct the
final datacube inside the MUSE pipeline, after having registered the astrometry 
to the available HST imaging with in-house IDL codes to achieve a sub-arcsecond precision
in both the absolute and relative astrometry. Data are interpolated over a regular grid of $0.2''$ 
spatially and $1.25$\AA\ spectrally. Finally, we apply the heliocentric correction to the wavelengths 
computed in air by the pipeline. An RGB color image produced by combining three slices of 1000\AA\ 
from the final MUSE datacube is shown in Figure \ref{fig:fov}. In this figure, 
we also superimpose a map of the \ha\ flux, obtained as described in the next section.  

The final datacube has an excellent image quality, with full width at half maximum (FWHM) 
of $0.73''\pm 0.05''$ for point sources in the field. We also characterize the noise properties by 
measuring in each spatial pixel the standard deviation of fluxes in the wavelength interval 
$6740-6810$\AA, which is clean of bright sky line residuals. The mean surface brightness limit
at $3\sigma$ confidence level is $3.0\times 10^{-18}$ erg~s$^{-1}$~cm$^{-2}$~\AA$^{-1}$~arcsec$^{-2}$
for Field A and $4.9\times 10^{-18}$ erg~s$^{-1}$~cm$^{-2}$~\AA$^{-1}$~arcsec$^{-2}$ for Field B.
Noise properties are uniform across the field, with a spatial variation expressed in 
unit of standard deviation about the mean of  
$4\times 10^{-19}$ erg~s$^{-1}$~cm$^{-2}$~\AA$^{-1}$~arcsec$^{-2}$ for Field A and 
$1\times 10^{-18}$ erg~s$^{-1}$~cm$^{-2}$~\AA$^{-1}$~arcsec$^{-2}$ for Field B.
For an unresolved line of $\Delta \lambda = 2.55$\AA\ at $\sim 6700$\AA, the line surface 
brightness limit becomes $7.6\times 10^{-18}$ erg~s$^{-1}$~cm$^{-2}$~arcsec$^{-2}$
for Field A and $1.2\times 10^{-17}$ erg~s$^{-1}$~cm$^{-2}$~arcsec$^{-2}$ for Field B.

\section{Emission and absorption line modeling}\label{sec:ana}

To characterize the kinematics of the gas in the disk and tail, and the galaxy 
stellar rotation, we extract maps of the zeroth, first and second moment of 
the \ha\  ($\lambda 6563$) line, together with a map of the first moment of 
the \ion{Ca}{II} triplet ($\lambda\lambda\lambda 8498,8542,8662$), 
as detailed below. Throughout this analysis, for consistency with previous work
\citep{sun10,jac14}, we assume a systemic heliocentric velocity of $v_{\rm sys} = 4661 \pm 46~$\kms\ for 
\eso, equivalent to redshift $z_{\rm sys}=0.01555 \pm 0.00015$.

\subsection{Emission lines}

Maps of the \ha\ line flux $F_{\rm H\alpha}$ (Figure \ref{fig:flux}),  radial velocity relative to systemic 
$\Delta v = v - v_{\rm sys}$ (Figure \ref{fig:vel}), and of the velocity dispersion $\sigma$ 
(Figure \ref{fig:sig}) are obtained from the final datacube using the IDL custom code {\sc kubeviz}.
With {\sc kubeviz}, we fit a combination of 1D Gaussian functions to the \ha\
and [NII] $\lambda\lambda 6548,6583$ lines, keeping the relative velocity separation 
and the flux ratio of the two [NII] lines constant. While measuring the intrinsic 
line width $\sigma$, we also fix the instrumental line width $\sigma_{\rm ins}$ at each wavelength
using an interpolation function constructed with a third degree polynomial fit to the resolution curve 
in the MUSE manual. At the observed wavelength of \ha\ computed for the systemic redshift of \eso\
($\lambda =6664.87~$\AA), we find $R=2611.9$ or $\sigma_{\rm ins}=49~$\kms, in agreement with the 
resolution measured from skylines within a resampled datacube before sky subtraction. 
The spatial variation of the spectral resolution is found to be negligible for emission lines 
with an intrinsic width of $\gtrsim 10~$\kms.

Before the fit, the datacube is median filtered in the spatial direction with a kernel 
of $5\times 5$ pixels to increase the individual $S/N$ per pixel. No spectral smoothing 
is performed. In each spectrum, the continuum  is then evaluated inside two symmetric windows 
around the \ha-[NII] line complex. Only values between the 40th and 60th percentile are modeled  
with a linear polynomial to estimate the continuum under each line. The Gaussian fit is performed on 
all the continuum-subtracted spaxels, assigning a minimum value of $\sigma=1~$\kms\ to spectrally 
unresolved emission lines. During the fit, {\sc kubeviz} takes into account the noise from the ``stat'' 
datacube, thus optimally suppressing skyline residuals. However, the adopted variance underestimates 
the real error, most notably because it does not account for correlated noise. We therefore renormalize 
the final errors on the fitted parameters so as to obtain a reduced $\tilde \chi^2\sim 1$. In the 
subsequent analysis, we mask spaxels where the $S/N$ of the \ha\ flux is $<3$ on the filtered data. 
Further masking is applied to the spaxels for which the line centroids or the line widths 
are extreme outliers compared to the median value of their distributions, or the errors on these quantities 
exceeds 50 \kms\ (i.e. conditions indicative of a non-converged fit). Throughout this analysis, the 
\ha\ flux has been corrected to account for appreciable Galactic extinction towards the Galactic plane 
\citep[$A_{\rm H\alpha}=0.55$;][]{sun07}, but not for internal extinction. However, our conclusions
primarily rely on kinematics that are not severely affected by the presence of internal dust.

As a consistency check on the velocity measurements, 
we also model the [SII] $\lambda\lambda 6717,6731$ doublet 
following the same procedure. The derived velocities $\Delta v$ from \ha\ and  [SII] are 
found to be in excellent agreement, to better than 1~\kms\ on average over the entire cube. 
The difference in the two reconstructed velocity maps is $\sim 18~$\kms\ for all regions where 
[SII] is detected at $S/N>3$, and $\sim 7~$\kms\ where $S/N>10$. A similar agreement is found 
when comparing the two measurements of velocity dispersion. In the following, we will 
only present and analyze the velocity field measured via \ha\ emission, the 
strongest transition, which we use as tracer for the bulk of the ionized hydrogen. A detailed analysis 
the properties of [SII] and other emission lines will be presented in the second paper of this series. 
We also note that, given the tracer used in our study, we will focus on the ram-pressure stripping of 
warm disk gas, which differ from strangulation, the stripping of the hot gas in the halo.

As a final check, we compare the fluxes of HII regions in our data with the 
values reported by \citet{sun07} in their table 1. We focus on 
three isolated regions \citep[ID 2, 3, and 7 in figure 1 of][]{sun07}, for which we
find a flux difference of 1\%, 3\%, and 6\%, respectively.
This test confirms the excellent agreement in the flux calibrations of the two data sets.

\subsection{Absorption lines}

\begin{figure}
\includegraphics[scale=0.55]{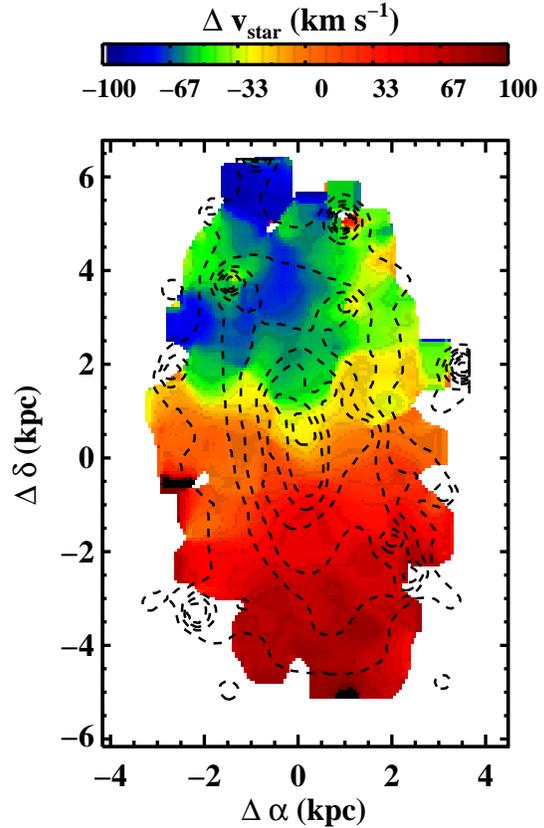}
\caption{Velocity map of the stellar component in the disk of \eso\ as traced
by the \ion{Ca}{II} triplet at a resolution of $60\times 60$ pc. 
The map has been smoothed with a median filter of  $10\times 10$ pixels for visualization 
purposes. The dashed contours mark isophotes measured on the stellar continuum between 
$\lambda = 8380-8950~$\AA, in steps of 0.4 mag~arcsec$^{-2}$. The image coordinate system 
is the same as the one adopted in Figure \ref{fig:flux}, but within a smaller box centered 
at the galaxy position. The ordered velocity field is aligned with 
respect to the stellar isophotes implying that gravitational interactions cannot be 
the dominant mechanism for gas removal.}\label{fig:stkin}
\end{figure}

Following a similar procedure to the one adopted for modeling emission lines, 
we extract the rotation curve of the stellar disk of \eso\ by fitting Gaussian functions to 
the \ion{Ca}{II} triplet. Specifically, we apply to the datacube 
a median filter of $10\times 10$ pixels in the spatial direction, with no filtering in the spectral 
direction. We then identify stars on the continuum-subtracted image in the rest-frame wavelength range 
$\lambda = 8380-8950~$\AA, which we masks with circular regions of $1.2''$. 
The velocity relative to systemic is then found by fitting all the spaxels 
which have $S/N>5$ per spaxels as measured on the filtered continuum image, using Gaussian functions 
with initial guesses at the rest-frame wavelengths of the \ion{Ca}{II} triplet lines. 
We find that the best result is achieved by fitting only the strongest  $\lambda8662$ line,
a choice which reduces the noise introduced by poorly converged fits on the weaker transitions
at the lowest $S/N$. A map of the reconstructed velocity field is in Figure \ref{fig:stkin}.

The use of Gaussian functions, although rather simple, is preferred to full 
template fitting when handling $\sim 10000$ spaxels in order to reduce the run time 
of the analysis by orders of magnitude. Nevertheless, we have conducted extensive tests 
comparing results from Gaussian fits to results from full template fitting on a coarsely binned datacube, 
so as to verify that our method yields accurate centroids. For this test, we rebin 
the data using the Voronoi tessellation method with the code developed by \citet{cap03}. 
The binning procedure is performed on a continuum image in the rest-frame 
wavelength range $\lambda = 8380-8950~$\AA, requiring a final $S/N$ in the 
re-binned regions of $>80$.  Galactic stars are again identified and masked.
We then fit to the optimally-combined spectra in each Voronoi cell templates from \citet{cen01}, 
using the Penalized Pixel-Fitting method (pPXF) developed by \citet{cap04}. During the fitting procedure, 
we assume an instrumental line width at relevant wavelengths of $\sigma_{\rm ins}=37~$\kms, computed as 
described in the previous section. Next, we apply our Gaussian fitting technique 
to the same binned spectra. When comparing these two independent measurements, we find a narrow 
Gaussian distribution, centered at $2~$\kms\ and with standard deviation of $15~$\kms, 
value which we assume as the typical uncertainty on the velocity measurement over the entire datacube.
As for the case of emission lines, higher signal-to-noise regions are most likely affected by smaller 
uncertainties. Furthermore, the 2D velocity field measured in the coarser Voronoi cells is 
in excellent agreement with the higher-resolution map recovered with Gaussian fits, 
confirming the robustness of the measurement presented in Figure \ref{fig:stkin}.

\begin{figure}
\includegraphics[scale=0.34, angle=90]{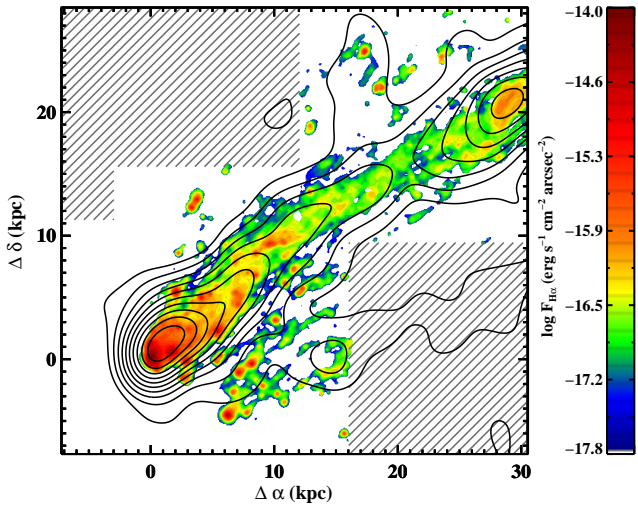}
\caption{Same as the left panel of Figure \ref{fig:flux}, but with logarithmic contours 
from Chandra observations in steps of 0.05 dex to show the location of the X-ray 
emitting gas. The X-ray map has been smoothed, but in fact the X-ray front is at the same position of the 
\ha\ front \citep[see][]{sun10}. The X-ray tails and the diffuse \ha\ emission appear to 
be co-spatial.}\label{fig:xray}
\end{figure}

\section{Discussion}\label{sec:disc}

Compared to traditional narrow band imaging in $80-100~$\AA\ wide filters, MUSE enables us
to extract from the datacube flux maps in very narrow wavelength windows 
($\sim 3-4~$\AA), in which we optimally weight the signal from emission lines over the 
background sky noise. In addition, by fitting emission lines in each spatial pixel, we can more precisely 
recover the line flux also in the presence of bright unrelated sources, such as foreground stars, if the 
Poisson noise introduced by these sources does not outweigh the signal. These key advantages, 
combined with the large VLT aperture, allow us to reconstruct a map of the 
\ha\ emission in the disk and tail of \eso\ which is one order of magnitude deeper than 
previous observations \citep{sun07}, down to surface brightness limits of
$F_{\rm H\alpha} \sim 10^{-18}$ erg~s$^{-1}$~cm$^{-2}$~arcsec$^{-2}$ (see Figure \ref{fig:flux}).

At this depth, our observations reveal that the \ha\ tail extends continuously for more 
than 30 kpc, with surface brightness 
$F_{\rm H\alpha} \sim 10^{-17}-10^{-15}$ erg~s$^{-1}$~cm$^{-2}$~arcsec$^{-2}$.
Figure \ref{fig:xray} also shows that the \ha\ emission is co-spatial in projection with the primary 
X-ray tail. At the west edge of our field of view 
($\Delta \alpha \sim 28~{\rm kpc},\Delta \delta \sim 21~{\rm kpc}$), one can see a significant surface
brightness enhancement, which is known to harbor $1.5-1.8\times 10^{8}$\msun\ 
of molecular gas \citep{jac14} and it is coincident with an X-ray bright region within the primary 
tail \citep{sun10}. On both sides of this primary tail, both towards north and south, we also detect 
a group of compact sources, some of which were previously identified as HII regions in 
spectroscopic observations by \citet{sun07} and \citet{sun10}. 

At the superior depth of our observations, the southern 
clump complex ($\Delta \alpha \sim 8~{\rm kpc},\Delta \delta \sim -2~{\rm kpc}$)
appears to be embedded in a lower surface brightness component, similar to the diffuse 
emission of the primary tail, stretching with the same projected orientation in the plane of the sky,
with a position angle of approximately $-54\deg$.  This second \ha\ tail is co-aligned with the 
secondary X-ray tail visible in the Chandra map, suggesting a common origin of the two gas phases,
which possibly originate from material that has been stripped from the southern part of the galaxy.
Unfortunately, our observations do not cover the full extent of this secondary tail, 
as visible for instance in figure 2 of \citet{sun10}. Additional observations in the south-west 
will thus be critical for a detailed investigation of the properties of the secondary tail.
Near the disk, embedded within the \ha\ emission of this secondary tail, there are few 
bright \ha\ knots, which \citet{sun10} identified as HII regions. 
In contrast to the more diffuse emission, these bright HII regions (and in particular the southernmost clump)
do not appear to be tightly correlated with the X-ray flux, hinting to a different physical 
origin of the emission in compact regions compared to the diffuse emission in the tails.

Our observations, in agreement with previous studies, point to ram-pressure stripping 
as the primary mechanism of gas removal. Indeed, we note how the stellar 
velocity field in Figure \ref{fig:stkin} exhibits an ordered velocity gradient along the major axis, 
as expected for a nearly edge-on rotating disk. Compared to the isophotes measured on an image 
extracted from the datacube in the wavelength range $\lambda = 8380-8950$\AA, the measured velocity 
field appears to be symmetric relative to the major axis and it is also well-centered at the optical 
location of \eso.  This ordered motion and an extremely disturbed morphology of the ISM compared 
to unperturbed isophotes support the conclusion of earlier studies that 
ram-pressure stripping is responsible for the gas ablation. Indeed, ram-pressure stripping produces 
only subtle transformations in the stellar disks \citep[e.g.][]{smi12b}, which largely retain memory 
of the original morphology and kinematics \citep[e.g.][]{bos08,tol11}. One may however argue that 
the velocity map in Figure \ref{fig:stkin} suggests the presence of small perturbations in the velocity 
field in the northern part of the stellar disk. However, we are more cautious in interpreting 
these features that are detected 
only in regions of low surface brightness and in proximity to two bright foreground stars, which 
may contaminate our measurement of the velocity fields at these locations.

Our observations, combined with the available HST imaging (see Figure \ref{fig:fov}), support a 
picture in which the galaxy suffered from ram pressure in the disk outskirts in the past, and it is 
still being ablated of its gas from the innermost regions.
Through models in which the efficiency of ram-pressure stripping depends on both gas density and 
galactocentric radius, as shown for example in figure 8 by \citet{jac14}, we can establish a 
correspondence between the spatial location of the \ha\ emitting gas and the time at which this 
gas was stripped. Moreover, lower density material in the less-bound outskirts of \eso\ is stripped 
more easily at earlier times. This means that material originating from the disk outskirts traces 
a more rapid stripping process which will come to an end before the stripping from the nucleus. 

Our map shows that the \ha\ bright knots that are offset from the primary
tail are being stripped from the northern and southern parts of the disk, respectively. 
At these large galactocentric radii, there is no evidence of \ha\ emitting gas inside the stellar disk 
of \eso\ to our sensitivity limit. Thus, ram pressure has already cleared the galaxy ISM in these 
less-bound regions. Conversely, the galaxy ISM is still present in the inner $2-3$ kpc of \eso\ 
\citep[see also][]{jac14}, and this material is still being stripped, feeding the primary tail.  
We therefore conclude that the primary tail originates from the inner regions of \eso\ and not 
from the northern trailing arm, as suggested by the earlier analysis of \ha\ narrow-band 
imaging \citep{sun10}. This outside-in stripping process offers a natural explanation for the different 
morphology of the primary \ha\ tail compared to the two complexes of \ha\ bright clumps in the north 
and south direction. We speculate that the diffuse gas from the outer disk was removed in the 
past and it has already mixed with the hot ambient medium downstream. As a consequence, we are now 
witnessing only the last episode of stripping of the densest clumps of gas, which trace the 
location of the older secondary tails.  This is also a plausible explanation for why the primary \ha\ tail 
is narrow compared to the optical disk \citep{sun10}. Originally, there may have been a 
single tail, fed by material stripped at all galactocentric radii, differently than what we see today.
A qualitative picture for how an inside-out stripping affects the morphology of the tail can be seen,
for instance, in the right panels of figure 1 in \citet{qui00}.

The analysis of the gas and stellar kinematics offers additional insight into how ram-pressure 
stripping occurs in \eso\ and how this galaxy moves as it approaches the
center of the cluster. The \ha\ map shows that \eso\ is moving towards the center the Norma 
cluster on a projected orbit with position angle of $\sim -54\deg$. The gas radial velocity, 
as traced by \ha\ (Figure \ref{fig:vel}), reveals no 
evident gradient along the primary tail, implying that the motion occurs in the plane of the sky.
Conversely, a clear velocity gradient is seen in the 
direction parallel to the galaxy major axis, perpendicular to the tail. This velocity pattern was 
already observed by \citet{sun10} who  studied selected HII regions with Gemini/GMOS spectroscopy. 
The unprecedented quality of the MUSE datacube, however, dramatically improves the fidelity 
with which the gas kinematics of \eso\ is traced, also within the diffuse component. 
Our observations confirm that \eso\ is approaching the cluster center on a nearly-radial orbit, with no 
significant tangential velocity component. The stripped gas retains a remarkable degree of coherence
in velocity to $\sim 20$ kpc downstream, 
which is originally imprinted by the rotation curve of the stellar disk.
The map of the stellar kinematics  (Figure \ref{fig:stkin}) shows a clear rotational 
signature in the plane 
of the sky, with peak to peak velocity of $\pm 70-80~$\kms\ in projection out to
radii of $4-5~$kpc. These values can be compared to the rotational velocity of $\sim 110-120~$\kms,
which \citet{jac14} infer from the $K-$band luminosity. 
Radial velocities in the range $\sim 80-120~$\kms\ are visible in the stripped gas primarily in the 
southern secondary tail, which in our picture originates from material that detached from the outermost 
part of the disk of \eso, at radii $\gtrsim 5~$kpc where stars rotate at 
comparable velocity \citep[see inset in Figure \ref{fig:vel} and cf. the rotational velocity estimates 
by][]{jac14}. A hint of the same effect, but with reversed 
sign, is visible in the northern part of the primary tail where the stripped gas is blue-shifted compared 
to systemic velocity, in agreement with the velocity field of the stellar disk.  

The infall velocity $v_{\rm inf}$ is unknown from direct observations,
but we can combine several pieces of evidence to conclude that \eso\
is fast moving towards the center of the cluster. 
Given the cluster velocity dispersion, which \citet{wou08} measured as 
$\sigma_{\rm cl} \sim 949$~\kms, one could derive an approximate 
estimate for the orbital velocity of \eso\ as $v_{\rm inf} \sim \sqrt3 \sigma_{\rm cl} \sim 1600~$\kms. 
However, the study of \citet{jac14} suggests that a much higher velocity $\gtrsim 3000$~\kms\ is 
needed for an orbit that crosses the cluster virial radius. 
Detailed hydrodynamic calculations are required to link the infall velocity to the 
disturbances seen in \eso, but here we simply note that such a high velocity is qualitatively 
in line with cases of extreme ram pressure. For \eso, gas is ablated 
even from the most bound regions of the galaxy potential well, implying high ram pressure.
Indeed, \citet{jac14} estimates that during its interaction 
with the dense ICM, with electron densities of $1.0-1.4\times 10^{-3}~\rm cm^{-3}$ \citep{sun10}, \eso\ 
can suffer from ram pressure up to a peak of $\sim 2.1\times 10^{-10}~\rm dyne~cm^{-2}$ for their 
assumed velocity.

As noted, a coherent radial velocity is visible on scales of at least 
$L_{\rm coe}\sim 20~$kpc from the galaxy center along the tail. 
Besides, the \ha\ emitting gas exhibits a modest 
degree of turbulent motion on similar scales, as seen from the velocity dispersion map
(Figure \ref{fig:sig}). The bulk of the gas has
a velocity dispersion $\sigma \sim 20-40~\rm$\kms, in qualitative agreement with the
narrow CO line profiles shown by \citet{jac14}. A lower velocity dispersion 
($\sigma \sim 15-25~\rm$\kms) is seen in the bright \ha\ knots which are 
offset from the primary tail, as expected for a colder medium which hosts sites of ongoing star formation.
Beyond $\sim 20~$kpc from the galaxy position, a higher degree of velocity 
dispersion can be seen, with typical values above $\sim 40~$\kms\ and peaks exceeding
$\sim 100~$\kms. We need to be more cautious in interpreting kinematics 
within Field B, due to the lower signal-to-noise of our observations. Nevertheless,
a progressively higher velocity dispersion moving far from the \eso\ disk along the tail
was also noted by \citet{jac14} who found larger FWHM and more complex line profiles in their
CO observations at larger distances from the disk. 
Moreover, Gaussian fits to coarsely binned data with higher $S/N$ 
show larger velocity widths in the outer tail, consistently with the measurements presented 
in Figure \ref{fig:sig}. 
For an infall velocity $\gtrsim 3000$~\kms, gas lying at $\gtrsim 20~$kpc 
from the \eso\ disk has been stripped no less than $\sim 6.5~$Myr ago,
assuming instantaneous acceleration of the removed material to the ICM velocity. 
Thus, our observations suggest that substantial turbulence due to fluid instabilities or residual rotational 
velocity in the stripped gas grows on timescales of at least $\sim 6.5~$Myr. However, both results of numerical 
simulations \citep[e.g.][]{roe08,ton12} and observations \citep{yos12,ken14} reveal that the stripped gas is not
accelerated to orbital velocity instantaneously, and that the denser medium does not
typically reach the infall velocity. Based on these results, the inferred timescale should 
be regarded as a conservative lower limit to the growth of instabilities in the tail.
It is also interesting to note that this laminar flow persists despite the shear between the ambient 
ICM and the \ha\ emitting, perhaps due to the presence of magnetic fields \citep{rus14,ton14}.  

Finally, we can appeal to the observed coherence to argue that the relative velocity $v_{\rm gas}$
between the gas tail and the galaxy likely exceeds the galaxy circular velocity $v_{\rm cir}$. 
Our argument starts by noting that the tail is confined by the ambient ICM.
If that were not the case, in the absence of a restoring centripetal force, material ablated from the disk 
at a radius $r_{\rm det}$ where the disk circular velocity is $v_{\rm cir} = v_{\rm los} / \sin i$ would leave 
on a tangential orbit, causing the tail to flare. Here, $i = 66\deg$ is the disk inclination angle and 
$v_{\rm los}$ is the observed line of sight velocity. Unsurprisingly, this is not what we observe, 
as the gas in the tail is confined by the ambient ICM pressure. 

In presence of a confining medium, a second velocity component perpendicular to the galaxy orbital motion 
contributes to the shear between the tail and the ICM. However, we noted how 
the tail retains a line-of-sight velocity similar to the disk rotation, which suggests 
that the tail cannot be spinning at higher velocity compared to the
velocity of the gas in the tail relative to the galaxy. In a simple geometry, a parcel of gas that leaves the 
disk with velocity $v_{\rm cir}$ rotates by an angle 
$\alpha_{\rm rot} \approx \frac{L_{\rm coe}}{r_{\rm det}} \times \frac{v_{\rm cir}}{v_{\rm gas}}$ 
by the time it has been displaced to a distance $L_{\rm coe}$ with velocity $v_{\rm gas}$.
Thus, the lack of substantial turbulence leads us to conclude that the angle $\alpha_{\rm rot}$
is likely small, which is to say that $v_{\rm cir} < v_{\rm gas}$ and there are no significant instabilities 
due to the disk rotation that grow substantially at the interface between the warm gas in the tail and 
the hot ICM. Given that $v_{\rm cir}\sim 80-100~\rm km~s^{-1}$ in the outer disk, and that most likely 
$v_{\rm gas} < v_{\rm inf}$, we also conclude that $v_{\rm cir} \ll v_{\rm inf}$, i.e. the galaxy is 
fast moving in the plane of the sky.

Clearly, this quite simple discussion should be verified with detailed 
hydrodynamic simulations designed to reproduce the properties of \eso, and that model the 
interaction of the warm/hot components, the effect of hydrodynamic and thermal instabilities, 
and the role of magnetic fields. All these quantities are in fact 
encoded in the phenomenology revealed by our MUSE observations.

\section{Summary and future work}

In this paper we have presented  MUSE science verification observations of the central disk and gaseous 
tail of \eso, a spiral galaxy infalling at high velocity on a nearly-radial orbit 
towards the center of the massive Norma cluster at $z \sim 0.0162$. During its encounter with the 
ICM, \eso\ experiences extreme ram-pressure stripping, which gives rise to
an extended tail that has been detected at multiple wavelengths, from the X-ray to the radio. 
The new capabilities of MUSE, with its large field of view, extended wavelength coverage, and high 
throughput, have allowed us to map the \ha\ emitting gas inside \eso\ and its tail to 
surface brightness limits of $F_{\rm H\alpha} \sim 10^{-18}$ erg~s$^{-1}$~cm$^{-2}$~arcsec$^{-2}$, 
and to reconstruct the stellar kinematics traced by the \ion{Ca}{II} triplet.

These new data offer an unprecedented view of \eso\ and its tail, which is traced in \ha\ 
emission to beyond 30 kpc.  These observations allow us to conclude that the ISM of
this galaxy suffered and it is still suffering from ram-pressure stripping at different 
galactocentric radii:  the less bound material has been completely stripped from the outer disk, 
giving rise to clumpy and irregular tails, while the galaxy center still contains part of the original 
ISM which feeds the primary \ha\ tail, aligned with the X-ray emission. The stripped gas retains 
the original rotation velocity of the stellar disk, but exhibits a remarkable degree of coherence in 
velocity space for $\sim 20~$kpc downstream. 
Beyond $\sim 20~$kpc from the galaxy disk along the tail, a greater degree of turbulence is seen, with peak 
velocity dispersion $\gtrsim 100~$\kms. These pieces of evidence are 
indicative that the galaxy is fast moving in the plane of sky, with infall velocity greater than 
the rotational velocity. For an infall velocity of $\sim 3000$~\kms, indicated by the orbital study 
of \citet{jac14}, material at $\sim 20~$kpc has been stripped no less than $\sim 6.5~$Myr ago, 
suggesting that the transition from laminar to turbulent flow occurs on comparable, or most likely, 
longer timescales. Finally, the ordered stellar rotation combined with regular isophotes in the disk
rules out gravitational interactions as the main process responsible for the stripping.

This work offers only a first glimpse of the rich MUSE datacube, in which we detect 
multiple transitions including [OIII], [OI], [NII], [SII], H$\beta$ in emission, as well as 
absorption lines within the stellar disk, most notably \ion{Ca}{II} and \ha. 
All these transitions will allow us to better constrain the physical properties of 
the disk and tail of \eso, including the properties of the bow shock and the 
interaction between the  stripped warm ISM and the hot X-ray emitting plasma.
Furthermore, a detailed comparison of these observations with results of numerical simulations
will be crucial to establish the hydrodynamic and thermal properties of the gas which is
subject to ram-pressure stripping, including the role of instabilities and magnetic fields. 
These studies will be presented in future papers of this series. 

Our work demonstrates the terrific potential that MUSE and future large field-of-view IFUs on 8-10m class 
telescopes have in studying ram-pressure stripping, building on 
earlier IFU studies \citep[e.g.][]{mer13}.
By targeting galaxies like \eso\ in which ram pressure is near to its maximum, these observations 
will provide key information on how gas is ablated from all galactocentric radii, how the diffuse 
and dense ISM components interact and mix with the ambient ICM, and on what timescales star formation 
responds to this gas removal.
Once applied to more typical cluster galaxies, results from these detailed analyses  
will offer a comprehensive view of how ram pressure affects the build-up 
of the observed red sequence in rich clusters. In turn, this will help to disentangle the role of 
environment from the role of internal processes in shaping the color bimodality seen in the 
densest regions of the Universe. 

\section*{Acknowledgments}

It is a pleasure to thank Mark Swinbank for his advice on how to handle MUSE data,
together with Emanuele Farina, David Wilman, and Elke Roediger for comments on this manuscript. 
We thank David J. Wilman and Joris Gerssen for the development of the KUBEVIZ code, and 
Eric Slezak for his contribution to the MUSE science verification programme. 
We thank the referee, Jeffrey Kenney,  for insightful comments on this manuscript.  
This work is based on observations made with ESO telescopes at the La Silla Paranal Observatory 
under programme ID 60.A-9349(A). M. Fumagalli acknowledges support by the Science and Technology 
Facilities Council [grant number  ST/L00075X/1]. M Fossati acknowledges the support of the 
Deutsche Forschungsgemeinschaft via Project ID 387/1-1. For access to the data used in this paper 
and the IDL code to process the MUSE datacubes, please contact the authors. 


\label{lastpage}

\end{document}